\documentclass[
    ,final            
  ]
  {aipproc}

\layoutstyle{8x11double}

\begin{document}

\title{Gamma-Ray Pulsar Studies with GLAST}


\author{D. J. Thompson}{
  address={Astrophysics Science Division, NASA Goddard Space Flight Center, Greenbelt, MD 20771 USA\\On behalf of the GLAST LAT Collaboration}
}



\begin{abstract}
 Some pulsars have their maximum observable energy output in the gamma-ray band, offering the possibility of using these high-energy photons as probes of the particle acceleration and interaction processes in pulsar magnetospheres. After an extended hiatus between satellite missions, the recently-launched AGILE mission and the upcoming Gamma-ray Large Area Space Telescope (GLAST) Large Area Telescope (LAT) will allow gamma-ray tests of the theoretical models developed based on past discoveries. With its greatly improved sensitivity, better angular resolution, and larger energy reach than older instruments, GLAST LAT should detect dozens to hundreds of new gamma-ray pulsars and measure luminosities, light curves, and phase-resolved spectra with unprecedented resolution.  It will also have the potential to find radio-quiet pulsars like Geminga, using blind search techniques.  Cooperation with radio and X-ray pulsar astronomers is an important aspect of the LAT team's planning for pulsar studies. 
\end{abstract}


\maketitle


\section{Gamma-ray Pulsars - Introduction}

  Balloon and small satellite observations revealed gamma radiation from the Crab (e.g.\cite{Browning71}) and Vela \cite{djt75} pulsars in the 1970's, becoming early examples of multiwavelength pulsar studies. The number of gamma-ray pulsars grew to at least seven during the Compton Gamma Ray Observatory (CGRO) mission in the 1990's (for a summary, see \cite{djt04}). These high-energy photons are produced by primary interactions of the energetic particles accelerated in the pulsar magnetosphere.  Their potential for studying some basic interaction processes was one stimulus for the development of pulsar models that included gamma rays as a characteristic.  With the recent launch of AGILE and the upcoming launch of GLAST, the gamma-ray window on pulsars offers new promise. 

\section{Challenges and Opportunities}

Any satellite observations of pulsars present difficulties not found with ground-based observatories.  Satellites move rapidly.  Clocks on satellites, particularly before the advent of GPS, require careful monitoring and calibration. Data streams are subject to transmission errors, as well as human errors in design and implementation.  Making corrections to a system in orbit is often non-trivial. 

For gamma-ray telescopes, the challenges are compounded by the low detection rate.  Even the brightest pulsars produce high-energy gamma rays separated by many pulse periods.  A weak pulsar like PSR B1055$-$52  yielded only 3 detected photons per day when in the field of view of EGRET on the Compton Observatory. During one day the pulsar rotates more then 400,000 times.  Gathering statistics required adding data from many observations, often spaced months or even years apart. 

The sparse nature of gamma-ray pulsar data has made the field highly dependent on timing information from other wavelengths, primarily radio.  The radio astronomy community has been highly supportive of the opposite end of the spectrum.  The GLAST team greatly appreciates the help of radio astronomers at Arecibo, GBT, Jodrell Bank, Nan{\c c}ay, Parkes, and other telescopes, as well as the X-ray timing programs being carried out with RXTE.  For further information about gamma-ray pulsar timing programs, see \cite{smith07}. 

\begin{figure}
  \includegraphics[height=.4\textheight]{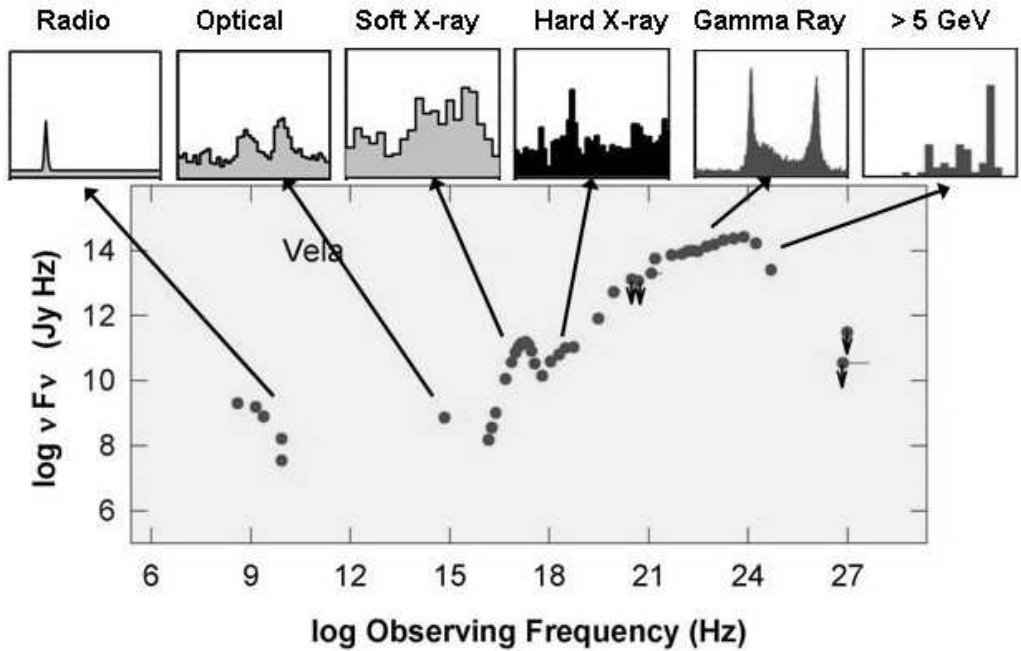}
  \caption{Multiwavelength observations of the Vela pulsar.  Top: light curves at various wavelengths. Bottom: spectral energy distribution of the pulsed radiation.  For references, see \cite{djt04}.}
\end{figure}

Offsetting the challenges of gamma-ray pulsar observations is the value of carrying out multiwavelength studies.  As in much of modern astrophysics, the synergy of broad-band research applies to pulsars. An example is Fig. 1, which shows multiple light curves and the pulsed spectral energy distribution for Vela, the brightest of the gamma-ray pulsars. Two features seem clear from this figure:

1. The pulsed energy output is overwhelmingly dominated by the high-energy emission, and the high-energy cutoff indicates that some sort of limit is reached in the GeV energy range;

2. The differences in the light curves across the spectrum illustrate multiple emission components, including coherent radio emission, thermal X-rays, and non-thermal gamma radiation. 

Trying to understand a multiwavelength object like this one by using only one energy band would be much like the proverbial blind men and the elephant.  They would all give incomplete and possibly-contradictory answers.  

The fairly small number of gamma-ray pulsars and their limited observational data have not solved the fundamental questions about high-energy particle acceleration and interactions in the magnetosphere.  Key questions that remain include:

\begin{itemize}

\item Where and how are the particles accelerated? Is it near the magnetic pole (polar cap), or in a slot gap or outer gap?

\item How and with what do the particles interact to produce gamma rays? 

\item Are the processes the same for all neutron star systems?

\item How does the complex environment (frame dragging, aberration, strong magnetic and electric fields, high currents) affect the observed radiation patterns?

\end{itemize}

One outcome of the variety of results seen for different high-energy pulsars has been a renewed interest in theoretical modeling.  Such models not only seek to explain existing observations but also to predict future patterns. The fact that models have been able to explain existing observations fairly well strengthens the need for better observations. As an example, Grenier and Harding \cite{iag06} summarize results showing that polar cap, slot gap, and outer gap models are capable of explaining, at least to first order, the double-pulse results for the Vela gamma-ray pulsar, given appropriate reasonable assumptions about the pulsar.
 
\section{New Gamma-Ray Satellite Telescopes}

\begin{figure}
  \includegraphics[height=.5\textheight]{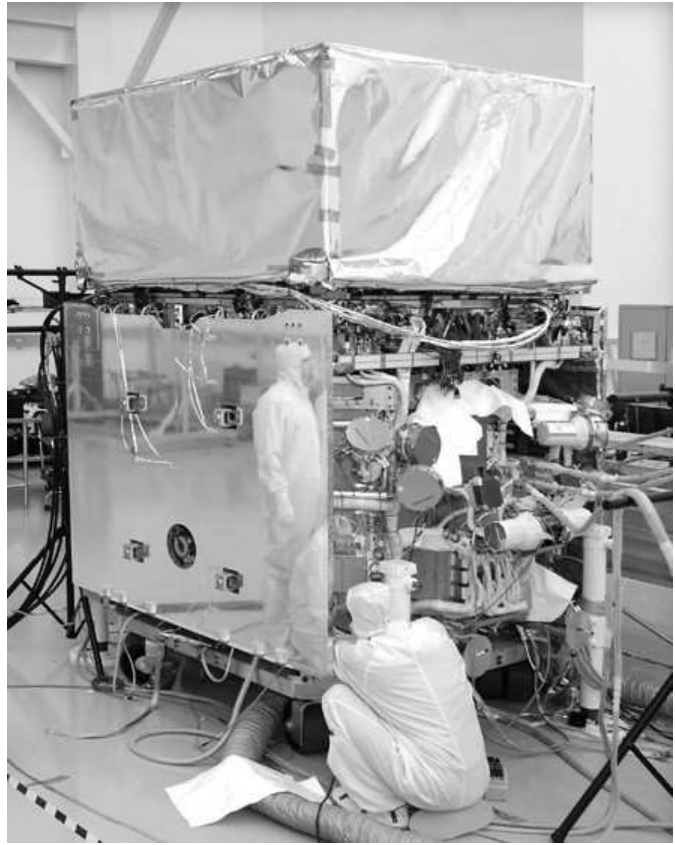}
  \caption{The GLAST Observatory.  The GLAST Burst Monitor detectors are located on the side of the spacecraft, to the lower right in the photo.  The Large Area Telescope is the box-like instrument on the top of the spacecraft. Photo: NASA and General Dynamics.}
\end{figure}

\subsection{AGILE}

AGILE (Astro-rivelatore Gamma a Immagini LEggero) is an Italian satellite launched on April 23 of this year \cite{MT06}. Their Web site, with current information about the mission, is {\url{http://agile.rm.iasf.cnr.it/ }}. AGILE is planned as a two-year mission. 

Like the earlier SAS-2, COS-B, and EGRET gamma-ray telescopes, AGILE relies on the pair production process for gamma-ray detection. It has a tracker to convert the gamma rays and determine the arrival direction (silicon strips instead of the older gas detectors), a calorimeter to measure energies (cesium iodide), and an anticoincidence detector to reject the huge background of charged particles in space (plastic scintillator).   

AGILE's high-energy detector operates in the energy range 30 MeV to 50 GeV.  It has a huge field of view (approximately 2.5 steradians) and therefore maps a large fraction of the sky for each pointing direction.  Although physically smaller than EGRET on the Compton Observatory, AGILE will have comparable source sensitivity and angular resolution. AGILE also has a thin, lightweight coded mask X-ray imager (called super-AGILE).  A critical parameter for pulsar studies is timing.  AGILE will have absolute time tags on individual gamma rays of a few microseconds. 

\subsection{GLAST}

GLAST (Gamma-ray Large Area Space Telescope) is currently in testing for launch in the Spring of 2008.  GLAST is designed as a major international facility with a minimum lifetime of five years (and no consumables that prevent a mission extending to 10 years or longer). The mission Web site is {\url{http://glast.gsfc.nasa.gov/}}.    The GLAST Observatory will carry two scientific instruments:

\begin{itemize}
\item The GBM (GLAST Burst Monitor) is a successor to BATSE on the Compton Observatory.  It will use a set of sodium iodide (NaI) and bismuth germanate (BGO) wide-field detectors to monitor the sky for transients in the 10 keV -- 30 MeV energy range  \cite{AV04}.  The GBM will be able to detect soft gamma repeaters, but it does not have a pulsar timing mode.

\item The LAT (Large Area Telescope) is the primary instrument on GLAST.  The LAT, which was called GLAST by itself in the early phases of the program, is a pair-production high-energy telescope successor to EGRET on CGRO.  It uses the same basic technology as AGILE (silicon strip tracker, CsI calorimeter, plastic scintillator anticoincidence detector) but on a much larger scale (Fig. 2) \cite{PM03}. For neutron star science, the LAT will be the principal GLAST instrument. 
\end{itemize}

\noindent Some important characteristics of the GLAST LAT are:

\begin{itemize}

\item Huge field of view ( approximately 2.4 steradians, or $\sim$20\% of the sky).

\item Planned scanning mode views the entire sky every 3 hours.

\item Broad energy range (20 MeV - $>$300 GeV, including the largely unexplored 10 - 100 GeV range).

\item Improved point spread function for gamma rays (a factor $>$3 better than EGRET for E $>$1 GeV).

\item Large effective area (factor $>$4 better than EGRET).

\item Single photon absolute time accuracy better than 10 microseconds.

\end{itemize}

This combination of improvements results in a factor $>$30 improvement in sensitivity compared to EGRET, with an even larger factor at energies above 10 GeV.

\subsection{Pulsar Science with the New Gamma-Ray Telescopes}

Pulsars will be primary science topics for both AGILE and GLAST.  Coupled with the rapid advance of pulsar astrophysics in other wavelength bands, the new gamma-ray capability offers many opportunities.  Some of these are:

\subsubsection{The EGRET Legacy of Pulsar Candidates}

The Compton Observatory mission ended in 2000, and the bulk of the observations with EGRET were done well before that time.  The EGRET data left behind some intriguing pulsar possibilities that the new telescopes will clearly resolve. 

\begin{itemize}
\item The EGRET measurements of pulsed emission from PSR B0656+14 \cite{PR96}, PSR B1046$-$58 \cite{VK00}, and PSR J0218+4232 \cite{LK00} were promising, although the confidence levels that the EGRET data were pulsed at the radio period were about 5 orders of magnitude lower than those of the best-known gamma-ray pulsars.  PSR J0218+4232 is of particular interest, because it is the only ms pulsar for which evidence is seen in gamma rays. The new instruments will have no difficulty confirming the nature of the gamma-ray emission from these sources. 

\item A significant number of young, energetic pulsars were found in EGRET unidentified source error boxes after the CGRO mission ended.  Such pulsars as PSR J1016$-$5857, PSR J1015$-$5719, PSR J1420$-$6048, PSR J1637$-$4642, PSR J1837$-$0559 and
PSR J2229+6114 have spin-down luminosities large enough that a few percent of their energy loss could power the gamma-ray emission \cite{MK03}, \cite{JH01}.  Due to timing uncertainties and the possibility of glitches, extrapolation back to the EGRET era requires too many trials to produce statistically-significant results \cite{DT02}. By detecting these, AGILE and GLAST could significantly expand the sample of gamma-ray pulsars. 

\end{itemize}

\subsubsection{GLAST LAT Searches for New Pulsars}

Although the final instrument response functions for the GLAST LAT are still in development, it is possible to estimate the LAT performance based on its design characteristics \cite{PM03}. To first order, detecting pulsed gamma radiation is limited by photon statistics (although pulse shape, diffuse gamma-ray backgrounds, spectral shape, and proximity of strong sources affect the ultimate performance). In two years with its scanning mode, LAT will detect 25-30 times as many photons for most pulsars as EGRET did in its lifetime. This improvement results in detections intrinsically 25 times fainter or 5 times farther away. Several of the known gamma-ray pulsars are at distances of 2 kpc; therefore LAT will be able to detect some pulsars at least as far away as the Galactic Center. 

The number of gamma-ray pulsars LAT may see depends not only on the LAT sensitivity, but also on the physics and beaming of the pulsars themselves.  Different models and analyses make predictions ranging from dozens to hundreds of new pulsars.  For a recent summary by Harding of some estimates and the implications of various results in terms of pulsar populations, see \cite{AH07}.  

One important caveat in reviewing any predictions of future gamma-ray pulsar detections is that LAT may be able to detect as sources more pulsars than it can confirm as pulsed sources.  In principle, a pulsed signal can be seen in gamma-ray data even when no source is visible in the spatial analysis (e.g. the EGRET detection of PSR B1951+32 \cite{PR95}), but such a pulsed detection requires good contemporaneous timing information from radio or other wavelengths.  If such timing data are not available, the LAT pulsed sensitivity is diminished due to the searching in period and period derivative space required.

The most extreme case is ''blind'' searching of gamma-ray data for pulsations. Even without the X-ray data \cite{JH92} that triggered the identification of Geminga as a gamma-ray pulsar, the pulsations could have been found in the EGRET data, and therefore the high sensitivity of LAT will enable effective periodicity searches of other gamma-ray sources \cite{AC01}. As noted by Ransom, a number of factors, including the timing noise that seems to be characteristic of gamma-ray pulsars and the scanning mode of GLAST, will make such searches very challenging \cite{SR07}.  Nevertheless, new searching techniques are being developed (e.g. \cite{MZ07}) to maximize the opportunities for finding such pulsars. It seems unlikely that Geminga is unique, and the chance to find a new population of pulsars warrants the effort.

\begin{figure}
  \includegraphics[height=.4\textheight]{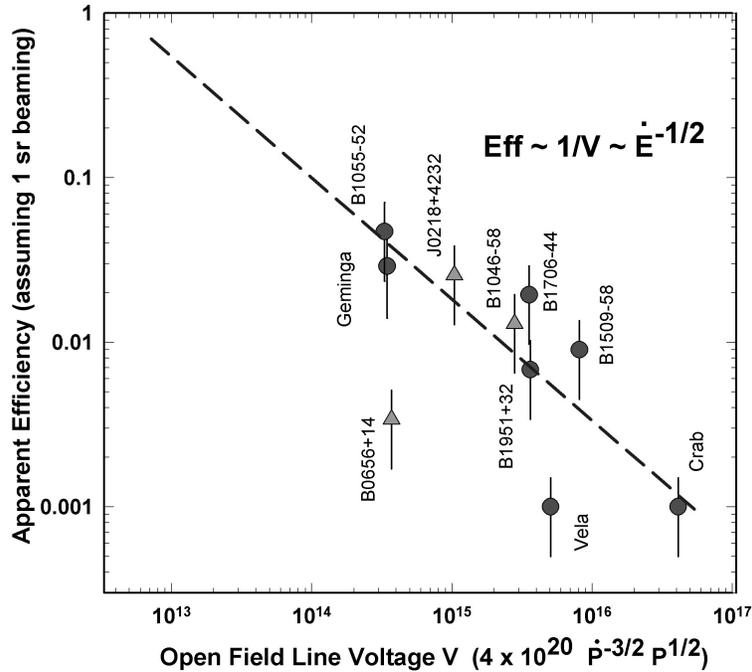}
  \caption{Efficiency (high-energy luminosity/spin-down luminosity) as a function of open field line voltage for high-confidence gamma-ray pulsars (circles) and lower-confidence gamma-ray pulsars (triangles).}
\end{figure}

\subsubsection{Testing Phenomenological Gamma-Ray Pulsar Properties}

Although the sample of gamma-ray pulsars is not extensive, it does provide some opportunities to search for trends in the data. Such phenomenological patterns will be obvious tests for the AGILE and GLAST gamma-ray pulsar populations. 

One trend, first noted by Arons \cite{JA96}, is that the efficiency for conversion of spin-down energy into high-energy radiation appears to be inversely proportional to the open field line voltage V.  A version of this pattern using the final EGRET measurements and recent distance estimates for these pulsars is shown in Fig. 3. Although it does not fit all the data, the relationship is reasonably good over two orders of magnitude. 

Some immediate questions that will be addressed by upcoming observations of a larger number of gamma-ray pulsars and greater detail of pulse shapes are:

\begin{itemize}
\item What happens at lower values of V?  The trend cannot continue, because the efficiency would reach 100\% somewhere below 10$^{13}$ volts.  Is there a sharp ''death line'' for gamma-ray emission, or a gradual roll-over? What is the lowest voltage for which gamma-ray emission is possible?

\item How dependent is this relationship on the (almost-surely incorrect) assumption that the gamma rays are all beamed into 1 steradian?  Gamma-ray pulses tend to be broad (Geminga, for example, radiates gamma rays essentially throughout the rotation of the neutron star). Do these pulse shapes imply a large beam or some preferential orientation? Are there major variations in the gamma-ray beam shape or size? 

\end{itemize}

\begin{figure}
  \includegraphics[height=.4\textheight]{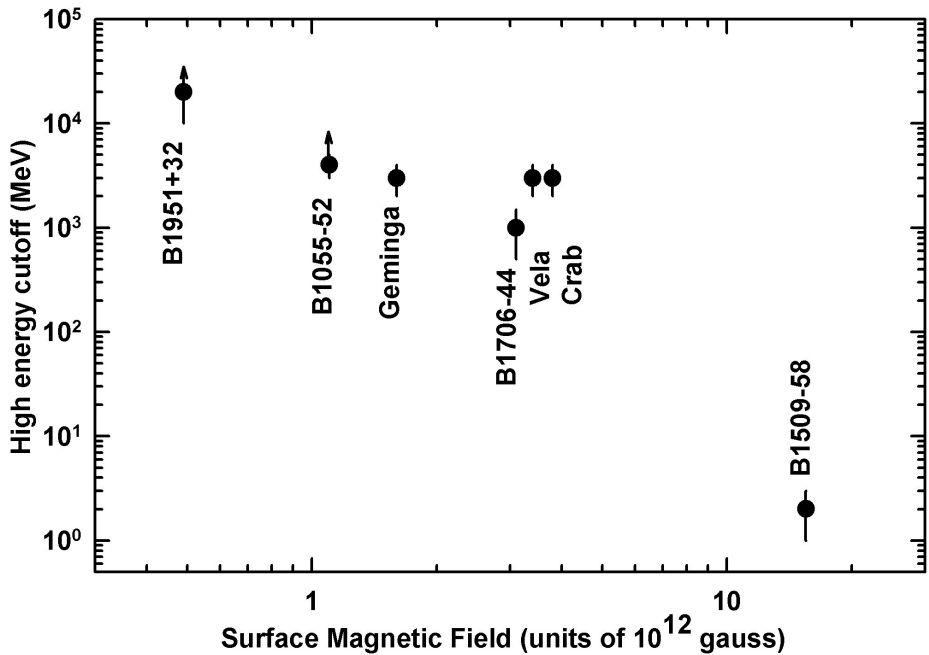}
  \caption{High-energy cutoff for gamma-ray pulsars as a function of surface magnetic field. }
\end{figure}

A feature observed with all gamma-ray pulsars is a high-energy cutoff.  Above some energy, the pulsed emission drops off dramatically. Fig. 4 shows a possible relationship between the approximate energy of this cutoff and the surface magnetic field of the pulsar, with a lower-energy cutoff being associated with a higher magnetic field.  The apparent pattern, though, is largely determined by two pulsars: PSR B1951+32, whose high-energy cutoff was not measured by EGRET but is implied by the lack of TeV emission; and PSR B1509$-$58, whose cutoff lies below the EGRET energy range. 

The new gamma-ray telescopes will test the hypothesis of this relationship in two ways:

\begin{itemize}
\item New gamma-ray pulsar detections will add statistics to this plot, increasing the sample.  

\item The GLAST LAT, with its greater sensitivity in the energy range above 10 GeV, will directly measure cutoffs for lower-field pulsars like PSR B1951+32 and PSR B1055$-$52, converting these lower limits into actual measurements. 

\end{itemize}

\subsubsection{Testing Gamma-Ray Pulsar Models}

Theory and modeling of high-energy pulsars have made significant advances in recent years, providing predictions that can be tested by future observations, then leading to better models. A full review of such work is beyond the scope of this paper, but some basic ideas are outlined here:

\begin{figure}
  \includegraphics[height=.4\textheight]{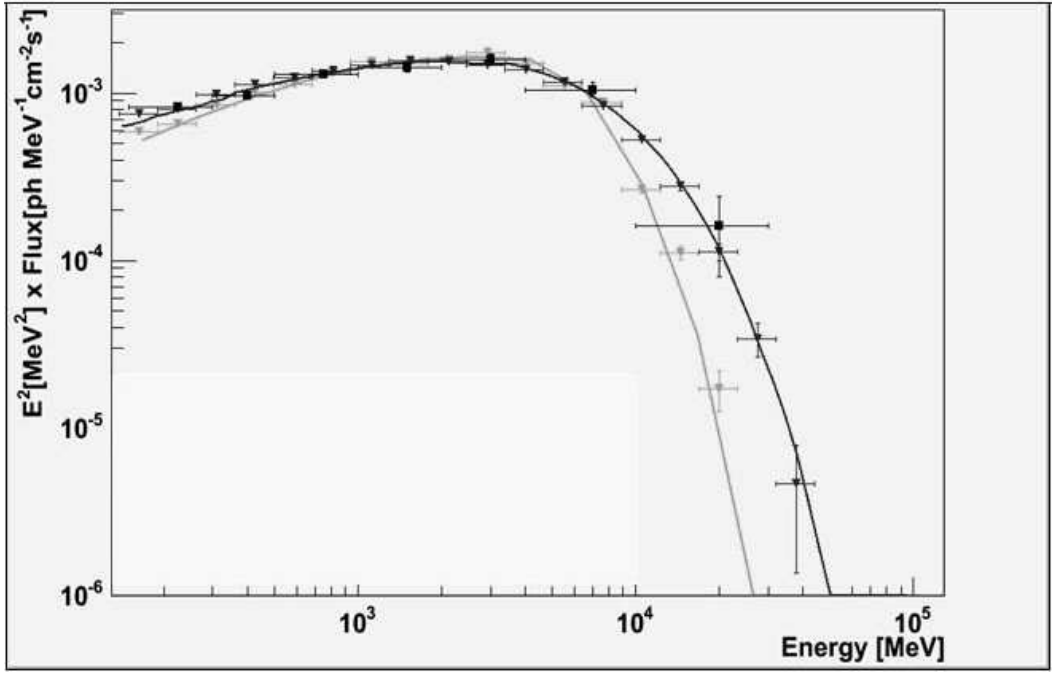}
  \caption{High-energy cutoff for the Vela pulsar, from \cite{MR07}. Black squares (with larger error bars): EGRET data; Black line and triangles: Outer gap model, with LAT one-year simulation; Gray line and triangles: Polar cap model, with LAT one-year simulation. }
\end{figure}

\begin{itemize}

\item Gamma-ray pulsar populations, both radio-loud and radio-quiet, offer important information about the location and geometry of the particle acceleration and interaction region in the pulsar magnetosphere. Romani and Yadigaroglu \cite{RR95} described this process in the context of an outer gap model. A recent study by Gonthier et al. \cite{PG07} includes a population synthesis using multiple pulsar models. Outer gap models typically predict more radio-quiet gamma-ray pulsars than do polar cap models. 

\item High-energy spectral cutoffs have been recognized as valuable discriminators between polar cap (which predict super-exponential cutoffs) and outer gap models (which predict exponential cutoffs). Modeling \cite{MR07} shows that LAT will be able to distinguish these cutoff shapes within a few months, and after one year will have a definitive measurement of the spectral shape at high energies. 

\item Phase-resolved spectra also differ significantly between pulsar models. Examples \cite{JD96}, \cite{RR96}, \cite{KC04} show that the EGRET spectra for even the brightest pulsars are not sufficiently well-defined to distinguish models.  With the much higher statistics and broader energy range of LAT, such distinctions should be visible. 

\item Millisecond pulsars are another opportunity to test theoretical modeling.  With only PSR J0218+4232 as a candidate ms gamma-ray pulsar \cite{LK00}, an important question is why nearby, bright PSR J0437$-$4715 was not seen \cite{JF95}, nor the ensemble of ms pulsars in 47 Tuc, for which EGRET found only an upper limit \cite{PM94}.  
Models (e.g. \cite{TB00}, \cite{ZC03}, \cite{AH05}) offer a variety of answers.  Most models predict a LAT detection of PSR J0437$-$4715, for example, but differ on whether other ms pulsars should be gamma-ray sources.

\end{itemize}

\subsection{Summary}
Gamma rays, with their direct relationship to primary acceleration and interaction processes, remain an important aspect of multiwavelength pulsar astrophysics.  AGILE provides a useful follow-on to EGRET on the Compton Observatory. The GLAST LAT will add significant capability for deeper and more detailed gamma-ray studies, allowing robust tests of both phenomenological and theoretical models of high-energy pulsar physics. 


\begin{theacknowledgments}
 I extend thanks to the members of the GLAST LAT Pulsar, Supernova Remnant, and Plerion Science Group for ongoing discussions of the LAT capability.  

\end{theacknowledgments}


\bibliographystyle{aipprocl} 




\end{document}